\documentclass[aps,preprint,a4paper,showpacs,showkeys,superscriptaddress]{revtex4-1}

\usepackage{latexsym}
\usepackage{amsmath,amssymb}
\usepackage{graphicx}
\usepackage{subfigure}
\usepackage{xcolor}
\usepackage{cancel}

\usepackage{hyperref}     
\hypersetup{colorlinks,%
  citecolor=blue,%
  linkcolor=cyan,%
  pdftex}

\begin{document}


\title{Unruh effect of nonlocal field theories with a minimal length}


\author{Yongwan Gim}%
\email[]{yongwan89@sogang.ac.kr}%
\affiliation{Department of Physics, Sogang University, Seoul, 04107,
  Republic of Korea}%
\affiliation{Research Institute for Basic Science, Sogang University,
  Seoul, 04107, Republic of Korea} %

\author{Hwajin Um}%
\email[]{um16@sogang.ac.kr}%
\affiliation{Department of Physics, Sogang University, Seoul, 04107,
  Republic of Korea}%

\author{Wontae Kim}%
\email[]{wtkim@sogang.ac.kr}%
\affiliation{Department of Physics, Sogang University, Seoul, 04107,
  Republic of Korea}%

\date{\today}

\begin{abstract}
The nonlocal field theory commonly requires a minimal length, and so it appears to
formulate the nonlocal theory in terms of
the doubly special relativity which makes the speed of light and the minimal length
invariant simultaneously.
We set up a generic nonlocal model having the same set of solutions as the local theory but allowing Lorentz violations due to the minimal length.
It is exactly corresponding to the model with the modified dispersion relation in
the doubly special relativity.
For this model, we calculate the modified Wightman function and the rate of response function
by using the Unruh-DeWitt detector method.
It turns out that the Unruh effect should be corrected by the minimal length
related to the nonlocality in the regime of the doubly special relativity.
However, for the Lorentz-invariant limit, it is shown that
the Wightman function and the Unruh effect remain the same as those of the local theory.
\end{abstract}

%



\maketitle


\section{Introduction}
\label{sec:intro}

There has been much attention to nonlocal theories
in light of a low energy description of fundamental nonlocal interactions.
Initially, nonlocality was mostly considered
in the context of axiomatic quantum field theory \cite{Efimov:1967pjn, Efimov:1968flw, Iofa:1969ex, Fainberg:1978cc, Fainberg:1992jt}.
And then many efforts have been devoted to studying various aspects of nonlocality in connection with gravity \cite{Tomboulis:1997gg, Moffat:2010bh,Biswas:2011ar,Isi:2013cxa,Briscese:2013lna,Calcagni:2014vxa,Frolov:2015bta}
and cosmology \cite{Deser:2007jk, Barnaby:2007ve, Barnaby:2008fk, Barnaby:2010kx, Deser:2013uya} as well as the role of nonlocality in the framework of string theory
\cite{Witten:1985cc, Brekke:1988dg, Eliezer:1989cr,  Ghoshal:2000dd,Moeller:2003gg, Ghoshal:2006te,Fuchs:2008cc, Calcagni:2013eua}.
In the nonlocal field theories,
 the nonlocality is commonly accompanied by a length scale $\ell$
because of the presence of higher derivative terms in the equations of motion.

There has been an intriguing issue on the Unruh effect
in a specific nonlocal field theory.
In the particular nonlocal theory
obeying the field equation of $\square e^{-\ell^2\square/2 } \phi=0$
with the minimal length $\ell$, the rate of response function
was calculated by using the Unruh-DeWitt detector method,
and then it was claimed
that there are significant modifications in the Unruh effect due to
the modification of the Feynman propagator originated from the nonlocality \cite{Nicolini:2009dr}.
However,
it was proved that, for a wide class of nonlocal theory obeying
the field equation given by $\square f(\ell^2 \square) \phi =0$ with the everywhere nonzero and analytic function $f$,
 the Bogoliubov coefficients should be exactly the same as
the case of a local theory
and thus
 the Unruh effect should remain unchanged
\cite{Kajuri:2017jmy}.
Recently,  for the specific nonlocal model of $\square e^{-\ell^2\square/2 } \phi=0$,
it was shown that the modified Feynman propagator due to the nonlocality consists of
the Wightman function along with the complementary error function
instead of the conventional step function \cite{Modesto:2017ycz}.
It means that the nonlocality is responsible for the error function
rather than the modification of the Wightman function,
so that the Unruh effect relying on the form of the Wightman function
naturally remains the same as
that of the local theory. 

On the other hand,
if the nonlocal theory should respect the special theory of relativity,
the length $\ell$ in the nonlocal theory
will be no longer minimal length due to the length contraction depending on the inertial frame.
So, the nonlocal field theory having the minimal length should be implemented by
 the doubly special relativity of the extended version of Einstein's special relativity  \cite{AmelinoCamelia:2000ge, AmelinoCamelia:2000mn},
  where the minimal length as well as the speed of light is an observer-independent scale.
In the framework of the doubly special relativity,
 the existence of the minimal length
would necessarily lead to
the modification of the dispersion relation such as
$f(\ell^2 k_0^2,\ell^2 k_i^2) k^\mu k_\mu =m^2$,
where $k^\mu k_\mu$ is related to the invariant speed of light and
$f(\ell^2 k_0^2,\ell^2 k_i^2)$ makes the minimal length $\ell$ an invariant scale
under any inertial frames \cite{Magueijo:2001cr, Magueijo:2002am, Kimberly:2003hp}.
If one were to consider the nonlocal model implemented by
the minimal length allowing the Lorentz violation, then
the field equation would be $\square f((i \ell \partial_0)^2,(i \ell \partial_i)^2) \phi= m^2$.
So, it would be interesting to study the Unruh effect for the nonlocal field theory
in the regime of the double special
relativity which makes the minimal length invariant.

In Sec.~\ref{sec:UDmethod},
 we will recapitulate the Unruh-DeWitt detector method for the Unruh effect in the local theory.
Then, in Sec.~\ref{sec:wightman} we will consider a massless nonlocal model of
$\square f((i \ell \partial_0)^2,(i \ell \partial_i)^2) \phi= 0$ which
respects the doubly special relativity, where
the function $f$ is an analytic function and everywhere nonzero.
Using the Unruh-DeWitt detector method presented in Sec.~\ref{sec:UDmethod},
we will obtain the modified Wightman function compatible with the doubly special relativity
and show that the Unruh effect receives some corrections due to the presence of
the minimal length in the nonlocal field theory.
If the nonlocal theory respects the Lorentz symmetry,
in other words, for the Lorentz-invariant limit of $\square f(\ell^2 \square) \phi =0$,
it turns out that the Unruh effect is the same as that of the local case as was shown
in Ref. \cite{Kajuri:2017jmy, Modesto:2017ycz}.
Finally, conclusion and discussion will be given in Sec.~\ref{sec:con}.

\section{Unruh-DeWitt detector method}
\label{sec:UDmethod}

We would like to encapsulate the Unruh-DeWitt detector method for the Unruh effect in the local field theory
presented in Ref.~\cite{ Unruh:1983ms}.
Let us consider a detector moving in Minkowski spacetime along a trajectory $x^\mu(\tau)$
with the proper time $\tau$,
and assume that it moves through a region permeated by a quantum scalar field $\phi(x)$.
The minimal interaction between the detector and the scalar field
 is given by the Lagrangian of $L_{\rm int}=g \mu(\tau) \phi\left(x(\tau)\right)$,
where $g$ is the small coupling constant and the detector operator $\mu(\tau)$  is approximated by
$\mu(\tau)=e^{i H_0 \tau} \mu(0) e^{-iH_0 \tau}$.
The detector will measure the energy transition
from the energy $E_0$ of the ground state to the energy $E$ of an excited state.
Since the coupling constant $g$ is small,
the transition probability is written as $P = \int dE |\mathcal{A}^{(1)}|^2$,
where the first order amplitude $\mathcal{A}^{(1)}$ is given by
\begin{align}
\mathcal{A}^{(1)} &= i\langle E;\psi | \int^{\tau}_{\tau_{\rm i}} L_{\rm int} d\tau |E_0; 0 \rangle \label{eq:Amp} \\
& = i g \langle E| \mu(0) |E_0 \rangle \int^{\tau}_{\tau_{\rm i}} d\tau e^{i \tau \Delta E} \langle \psi | \phi(x) |0 \rangle.
\end{align}
Note that $|0\rangle$ is the Minkowski vacuum and $|\psi \rangle$ is the excited state,
and the energy difference between them is denoted by $\Delta E = E-E_0$.
Then, the transition probability $P$ is obtained as
\begin{equation}
P = g^2 \int dE |\langle E| \mu(0) |E_0 \rangle|^2 R(\Delta E),
\end{equation}
where the response function $R(\Delta E)$ is written as
\begin{align}
R(\Delta E) &= \int^{\tau'}_{\tau'_{\rm i}} d\tau' \int^{\tau}_{\tau_{\rm i}} d\tau e^{-i ( \tau-\tau') \Delta E} \langle 0 | \phi \left(x(\tau)\right) \phi\left(x(\tau')\right) |0 \rangle \\
&= \int^{\Delta \tau^+}_{\Delta \tau^+_{\rm i}} d\Delta \tau^+ \int^{\Delta \tau^-}_{\Delta \tau^-_{\rm i}} d\Delta \tau^- e^{-i \Delta \tau^- \Delta E}  G^+(\Delta \tau^+,\Delta \tau^-)
\end{align}
with $\Delta \tau^\pm = \tau \pm \tau'$, and the positive frequency
Wightman function $G^+$ is defined as
\begin{equation}\label{eq:G+}
G^+(\Delta \tau^+,\Delta \tau^-)= \langle 0 | \phi \left(x(\tau)\right) \phi\left(x(\tau')\right) |0 \rangle.
\end{equation}
So, one can obtain the transition probability per unit time as
\begin{equation}
\dot{P} = g^2 \int dE |\langle E| \mu(0) |E_0 \rangle|^2 \dot{R}(\Delta E)
\end{equation}
with the rate of the response function 
\begin{align}\label{eq:Rdot}
\dot{R}(\Delta E) =  \int^\infty_{-\infty} d\Delta \tau^- e^{-i \Delta \tau^- \Delta E}  G^+(\Delta \tau^+,\Delta \tau^-),
\end{align}
where the integration range of $\Delta \tau^- $ is extended up to $\pm \infty$.

In the local field theory, the field equation of the free scalar field $\phi$ is given by
\begin{equation}\label{eq:EOMlocal}
\square \phi=0,
\end{equation}
and the  solution is written as
\begin{equation}\label{eq:sol1}
\phi(x)=\int \frac{d^3 k}{(2\pi)^3 2 \omega} \left[a_k e^{ik_\mu x^\mu}+a^\dag_k e^{-ik_\mu x^\mu}\right],
\end{equation}
 where the coefficients $a_k$ and $a^\dag_k$ are operators in such a way that
 the Minkowski vacuum is annihilated by the operator $a_k$,
 {\it i.e.}, $a_k  |0\rangle =0$.
By imposing the equal-time commutation relations,
$[\phi(\vec{x},t),\pi(\vec{y},t)]=i\delta^{(3)}(\vec{x}-\vec{y})$ and $[\phi(\vec{x},t),\phi(\vec{y},t)]=[\pi(\vec{x},t), \pi(\vec{y},t)]=0$
with the conjugate momentum $\pi=\dot{\phi}$,
we obtain the following quantization rules,
\begin{align}
[a_k, a^\dag_{k'}]=(2 \pi)^3 2\omega \delta^{(3)}(\vec{k}-\vec{k}'),\qquad [a_k, a_{k'}]=[a^\dag_k, a^\dag_{k'}]=0.  \label{eq:aat}
\end{align}
Plugging Eqs.~\eqref{eq:sol1} and \eqref{eq:aat}
into Eq.~\eqref{eq:G+},
one can also obtain the positive frequency Wightman function in the local field theory as
\begin{align}
G^+(x,x') &= \int \frac{d^3 k d^3 k'}{(2 \pi)^6 2\omega 2\omega '} \langle 0| [a_k, a^\dag_{k'}] |0 \rangle e^{i(k_\mu x^\mu-k'_\mu x'^\mu)} \\
&= \int \frac{d^3 k }{(2 \pi)^3 2\omega }  e^{i k_\mu (x^\mu-x'^\mu)}. \label{eq:G+local}
\end{align}

In the Minkowski spacetime with the coordinates of $x^\mu=(t,x,y,z)$,
the hyperbolic trajectory describes a uniformly accelerated detector along the $x$-axis
with the proper acceleration $a$ and the proper time $\tau$,
which is given by
\begin{equation}\label{eq:Rindler}
t(\tau)=\frac{1}{a} \sinh(a\tau),\qquad x(\tau)=\frac{1}{a} \cosh(a\tau)
\end{equation}
with the fixed $y$ and $z$.
By using Eqs.~\eqref{eq:Rdot} and ~\eqref{eq:G+local} in the hyperbolic trajectory \eqref{eq:Rindler},
the rate of response function of the uniformly accelerated detector can be calculated as
\begin{equation}\label{eq:Rdotlocal}
\dot{R}(\Delta E)=\frac{\Delta E}{2\pi}\frac{1}{e^{\frac{2\pi \Delta E}{a}}-1}.
\end{equation}
Eventually, the temperature can be  read off
from Eq.~\eqref{eq:Rdotlocal}
 as
\begin{align}
T_{\rm loc} =\frac{a}{2\pi}, \label{eq:T_U}
\end{align}
which is the well-known Unruh effect in the local field theory \cite{Unruh:1983ms}.
In the next section, we shall discuss the Unruh effect in the nonlocal field theory
in order to figure out
 how the conventional Unruh effect is modified by the nonlocality.

\section{Modified Wightman function and Unruh effect}
\label{sec:wightman}

There appears a minimal length in nonlocal formulation of theories with
higher derivatives, so that
the special relativity related to the Lorentz symmetry
could be promoted to the doubly special relativity where
the minimal length as well as the speed of light is invariant in any inertial frames.
Let us introduce a nonlocal model where the minimal length $\ell$ is invariant
under any inertial frames allowing the
conventional Lorentz violation \cite{Kimberly:2003hp},
which is generically described by the field equation of
\begin{equation}\label{eq:EOM1}
\square f\left((i\ell\partial_0)^2,(i \ell \partial_i)^2 \right) \phi_{\rm NL}=0
\end{equation}
with $\partial_i^2= \partial_x^2+\partial_y^2+\partial_z^2$,
where the function $f$ is an analytic function and everywhere nonzero.

 Note that
 one might worry about the presence of ghosts in the nonlocal model \eqref{eq:EOM1}
 since the addition of higher derivative terms
 could cause ghost-like excitations. Fortunately,
the pole structure of the nonlocal model \eqref{eq:EOM1} does not change thanks to the assumption that the function $f$ is analytic and everywhere nonzero, so that the ghost problem would be avoided.
 However, the present work does not contain a proper constraint
analysis, which is necessary to fully establish the stability of the quantum version of the
theory. So, one should
 perform the Hamiltonian analysis for nonlocal theories of infinite order by using the formalism of Refs.~\cite{Gomis:2000gy,Gomis:2003xv}, which allows a more
transparent identification of the physical phase space of an infinite derivative theory and reduces to the Ostrogradski construction in the finite derivative case \cite{Barnaby:2007ve}.
 



Additionally, we note that the dispersion relation corresponding to Eq.~\eqref{eq:EOM1} is
written as $f(\ell^2 k_0^2,\ell^2 k_i^2) k^\mu k_\mu =0$,
where $k^\mu k_\mu$ gives the invariant speed of light and
$f(\ell^2 k_0^2,\ell^2 k_i^2)$ makes the minimal length $\ell$ an invariant scale
under any inertial frames \cite{Magueijo:2001cr, Magueijo:2002am, Kimberly:2003hp}.
Also, we assumed that the function $f $ is nonzero,
so that the frequency $\omega$ satisfies $\omega=|\vec{k}|$, which
means that the speed of light is still invariant in spite of
the modified dispersion relation in the light of the doubly special relativity.

According to Ref.~\cite{Barnaby:2007ve},
the number of independent solutions to an infinite order differential equation is equal to the number of poles in its propagator.
Since the number of poles for the propagator of the nonlocal model \eqref{eq:EOM1} is the same as that of the local theory,
the complete set of solutions to the nonlocal field equation \eqref{eq:EOM1} is just that to the field equation \eqref{eq:EOMlocal} of the local theory.
 So, combining the plane wave solution of $e^{\pm i k_\mu x^\mu}$ as the set of the solutions to Eq.~\eqref{eq:EOM1}
gives
the field expansion as
\begin{equation}\label{eq:sol2}
\phi_{\rm NL}(x)=\int \frac{d^3 k}{(2\pi)^3 2 \omega} \left[b_{k} e^{i k_\mu x^\mu}+b^\dag_{k} e^{-i k_\mu x^\mu}\right].
\end{equation}

Now, one might wonder where the nonlocal effects including the Lorentz violation are reflected in the field expansion \eqref{eq:sol2}.
In the case of the nonlocal model \eqref{eq:EOM1}, it has the same set of the solutions as that of
the local theory,
so that nonlocal effects are not in the plane wave solutions
but in the coefficients $b_k$ and $b^\dag_k$,
and the commutation relations between them would be modified.

 Here,
one can consider the action for the equation of motion~\eqref{eq:EOM1} which is written as
 \begin{equation}\notag
S_{\rm NL}=\int d^4x \sqrt{-g} \left(-\frac{1}{2}\right) \phi_{\rm NL}(x)\left[-\square f\left((i\ell\partial_0)^2,(i \ell \partial_i)^2 \right)\right] \phi_{\rm NL}(x),
\end{equation}
and then the canonical momentum is obtained as $\pi_{\rm NL} = \delta S_{\rm NL}/\delta \dot{\phi}_{\rm NL} = -\sqrt{-g}g^{0\nu}\partial_\nu f \phi_{\rm NL}$.
However, even if the canonical momentum can be defined, the nonlocality does not allow us to convert the relation between the velocity  $\dot{\phi}_{\rm NL}$ and the canonical momentum $\pi_{\rm NL}$. 
So, it seems to be difficult to construct a Hamiltonian naively from the Lagrangian by a Legendre transformation.

 Let us see how the nonlocal effects are taken into account
in the coefficients $b_k$ and $b^\dag_k$ and what the commutation relations are.
 So, we redefine the field as
$\phi_{\rm NL} = f^{-1}\left((i\ell\partial_0)^2,(i\ell\partial_i)^2\right) \phi$
to obtain the commutation relation between the operators $b_k$ and $b^\dag_k$,
and thus the field equation \eqref{eq:EOM1} can be rewritten as $\square \phi = 0$.
Now the nonlocal field $\phi_{\rm NL}$ is written in terms of the local field $\phi$
by using Eq.~\eqref{eq:sol1},
\begin{align}
\phi_{\rm NL}(x) &=\frac{1}{f((i\ell\partial_0)^2,(i\ell\partial_i)^2)} \phi(x)  \notag \\
&=\int \frac{d^3 k}{(2\pi)^3 2 \omega} \left[\frac{a_k}{f(\ell^2 \omega^2, \ell^2 \vec{k}^2)} e^{ik_\mu x^\mu}+\frac{a^\dag_k}{f(\ell^2\omega^2,\ell^2 \vec{k}^2)} e^{-ik_\mu x^\mu}\right] \label{eq:sol3}.
\end{align}
By comparing Eq.~\eqref{eq:sol3} with the field expansion \eqref{eq:sol2},
the relations between the local operators of $a_k$ and nonlocal operators of $b_k$ are obtained as
\begin{equation}
b_{k} =\frac{a_k }{f(\ell^2\omega^2,\ell^2 \vec{k}^2)},\qquad b^\dag_{k} =\frac{a^\dag_k}{ f(\ell^2\omega^2,\ell^2 \vec{k}^2)}.
\end{equation}
From the commutation relation \eqref{eq:aat} between $a_k$ and $a^\dag_{k'}$,
we can obtain a relation between $b_k$ and $b^\dag_{k'}$ as
 \begin{equation}\label{eq:bbt}
[b_k, b^\dag_{k'}]=\frac{(2 \pi)^3 2\omega \delta^3(\vec{k}-\vec{k}')}{f(\ell^2\omega^2,\ell^2 \vec{k}^2)f(\ell^2\omega^2,\ell^2 \vec{k}'^2)},
\end{equation}
 where the vacuum of the nonlocal field theory is characterized by  $b_k |0 \rangle =0$.

Next, from the definition of the positive frequency Wightman function \eqref{eq:G+},
we can find the modified Wightman function corresponding to Eq.~\eqref{eq:EOM1}  as
\begin{align}
G^+(x,x') &= \int \frac{d^3 k d^3 k'}{(2 \pi)^6 2\omega 2\omega '} \langle 0| [b_k, b^\dag_{k'}] |0 \rangle e^{i(k_\mu x^\mu-k'_\mu x'^\mu)} \notag \\
&= \int \frac{d^3 k}{(2 \pi)^3 2\omega} \frac{1}{f^2(\ell^2\omega^2,\ell^2 \vec{k}^2)} e^{i k_\mu( x^\mu- x'^\mu)}\label{eq:G+nonlocal}.
\end{align}
It is worth noting that
in the field expansion \eqref{eq:sol3}
the modifications due to the nonlocality can be subsumed into the mode functions of
$\mathcal{F}_{k,\pm}^{\rm NL}(x)=[(2\pi)^3 2 \omega_k]^{-1/2} f^{-1} e^{\pm i k x}$
with the standard commutation relations \eqref{eq:aat}
instead of  the standard mode functions
of $\mathcal{F}_{k,\pm}(x)=[(2\pi)^3 2 \omega_k]^{-1/2} e^{\pm i k x}$
 with the modified commutation relations \eqref{eq:bbt}.
Interestingly,
the Wightman function is always written as Eq.~\eqref{eq:G+nonlocal}
regardless of whether the nonlocality is included in the mode functions or the operators.


Since $f$ is everywhere nonzero and analytic function,
it can be represented by a power series
with the help of $\omega=|\vec{k}|$,
\begin{equation}\label{eq:series}
\frac{1}{f^2(\ell^2\omega^2,\ell^2 \vec{k}^2)} = \frac{1}{f^2(\ell^2\omega^2,\ell^2 \omega^2)}
=\sum^\infty_{n=0} \alpha_n (\ell^2 \omega^2)^{n},
\end{equation}
where the coefficient $\alpha_0$ is fixed by $\alpha_0 =1$
to recover the local case when the nonlocal effects disappear.
Then, the positive frequency Wightman function \eqref{eq:G+nonlocal} is
explicitly calculated as
\begin{align}
G^+(x,x')
=\sum^\infty_{n=0} \alpha_n \ell^{2n} \frac{(-1)^n (2n)!}{8\pi^2 \Delta x} \frac{(\Delta t +\Delta x)^{2n+1}-(\Delta t -\Delta x)^{2n+1}}{(-\Delta t^2 +\Delta x^2)^{2n+1}}, \label{eq:G+xx}
\end{align}
where $\Delta t = t-t'$ and $\Delta x = |\vec{x}-\vec{x}'|$.
When the detector follows the hyperbolic trajectory \eqref{eq:Rindler},
the modified Wightman function  is obtained as
\begin{equation}\label{eq:G+nolocalR}
G^+(\tau,\tau')= \sum^\infty_{n=0} \sum^{2n}_{m=0} \alpha_n \ell^{2n} \frac{(-1)^{n+1}(2n)!}{4\pi^2} \frac{e^{a(n-m)\Delta \tau^+}}{\left(\frac{2}{a}\sinh\left(\frac{a}{2}\Delta \tau^-\right)\right)^{2(n+1)}}
\end{equation}
with $\Delta \tau^\pm = \tau \pm \tau'$.
So,
the rate of response function in the nonlocal model \eqref{eq:EOM1} is calculated as
\begin{equation}\label{eq:finaldotR}
\dot{R}(\Delta E)=\frac{\Delta E}{2\pi}\frac{1}{e^{\frac{2\pi \Delta E}{a}}-1}\left[1 +\sum^\infty_{n=1} \sum^{2n}_{m=0} \frac{\alpha_n \ell^{2n}}{2n+1} e^{a(n-m)\Delta \tau_+} \prod^n_{s=1}(s^2 a^2 + \Delta E^2) \right],
\end{equation}
where it reduces to the standard local limit for $\ell \to 0$.
The similar rate of response function can be found
in the Lorentz-violating cases \cite{Rinaldi:2008qt, Gutti:2010nv, MartinMartinez:2012th, Harikumar:2012yu, Harikumar:2012ff, Majhi:2013koa}.
Consequently,
in the generic nonlocal model \eqref{eq:EOM1},
the Wightman function and the rate of response function were corrected by the minimal length related to the nonlocality,
so that
the Unruh effect could be modified.

For example,
let us see the Unruh effect for a nonlocal model
defined by the doubly special relativity in Ref.~\cite{Kimberly:2003hp},
 which is given by
\begin{equation}\label{eq:toy1}
\frac{\square}{1-\ell^2 \partial_0^2} \phi_{\rm NL} =0,
\end{equation}
where the nonlocal term is $f=(1+\ell^2 \omega^2)^{-1}$ in the momentum space
and thus $f^{-2} =1+2\ell^2 \omega^2+\ell^4 \omega^4$.
So, one can easily
identify the coefficients $\alpha_n$ related to the nonlocality in Eq.~\eqref{eq:finaldotR} as
$\alpha_0=1$, $\alpha_1 =2 $, $\alpha_2 = 1$ and $\alpha_{n>2}=0$,
and thus the rate of response function is obtained as
\begin{align}
\dot{R}(\Delta E) = \frac{\Delta E}{2\pi}\frac{1}{e^{2\pi \frac{\Delta E}{a}}-1}&\left[1 + \frac{2\ell^2a^2}{3} A_1(a,\Delta \tau_+)\left( \frac{ \Delta E^2}{a^2}+1\right)  \right. \notag \\
&+ \left. \frac{\ell^4a^4}{5}A_2(a,\Delta \tau_+)\left(\frac{\Delta E^2}{a^2}+1\right)\left(\frac{\Delta E^2}{a^2}+4\right)  \right], \label{eq:dotR_toymodel}
\end{align}
with $A_1(a,\Delta \tau_+)=1+ 2\cosh(a\Delta \tau_+)$,
$A_2(a,\Delta \tau_+)= 1+ 2\cosh(a\Delta \tau_+)+2\cosh(2 a\Delta \tau_+)$.
Here, we note that Eq.~\eqref{eq:dotR_toymodel} depends on the proper time
and
such time-dependent rate of response functions also appear in
the case of superluminal dispersion relations \cite{Rinaldi:2008qt}
and $\kappa$-deformations \cite{Harikumar:2012yu, Harikumar:2012ff}.
The dependency of the proper time
means that the system is not in a global thermodynamic equilibrium.
So,
we take the approximation as $a \ll \Delta E $ to make Eq.~\eqref{eq:dotR_toymodel}
as a slight deviation out of the global thermodynamic equilibrium,
 and thus we assume a local thermodynamic equilibrium.
Finally, we can read off the Unruh temperature
by using the relation between the Planck distribution and the rate of response function of $\dot{R}(\Delta E)=(\Delta E/2\pi)(e^{\Delta E/T}-1)^{-1}$ \cite{Unruh:1983ms}
as
\begin{align}
T= \frac{\Delta E}{\ln\left(1+ \frac{\Delta E}{2\pi \dot{R}(\Delta E)} \right)}. \label{eq:T1}
\end{align}
This procedure is very similar to derivation of the Hawking temperature from the scattering amplitude \cite{Ghoroku:1994np,Natsuume:1996wf, Kim:2002nc, Clement:2007tw}.
Then, the temperature can be read off from Eq.~\eqref{eq:T1} as
 \begin{equation}\label{eq:tem}
T_{\rm NL} = \frac{a}{2\pi} + \frac{a^2}{2\pi \Delta E}\ln \left(1+\frac{2 \ell^2 \Delta E^2}{3}A_1(a,\Delta \tau_+)+\frac{\ell^4 \Delta E^4}{5}A_2(a,\Delta \tau_+)\right),
\end{equation}
which shows that the Unruh effect could be corrected by the nonlocality.

Next, let us see the Unruh effect for
a wide class of the nonlocal model
 respecting the Lorentz symmetry such as
\begin{equation}\label{eq:kajuri}
\square f(\ell^{2}\square) \phi_{\rm NL}= 0,
\end{equation}
which was already considered by using the Bogoliubov transformation in Ref.~\cite{Kajuri:2017jmy}.
By replacing $f(\ell^{2}\omega^2,\ell^{2}\vec{k}^2)$ with $f(\ell^{2}k^\mu k_\mu)$
in Eqs.~\eqref{eq:EOM1} and~\eqref{eq:G+nonlocal},
the positive frequency Wightman function is modified
 as
\begin{align}\label{rel}
G^+(x,x')= \int \frac{d^3 k}{(2 \pi)^3 2\omega}
\frac{1}{f^2(\ell^{2} k^\mu k_\mu)} e^{i k_\mu( x^\mu- x'^\mu)}.
\end{align}
Note that the nonlocal effect actually disappears because
$f(\ell^{2} k^\mu k_\mu)=f(0)=1$ due to the fact that $k^\mu k_\mu=-\omega^2+\vec{k}^2=0$,
which in essence amounts to the limit of $\ell \rightarrow 0$.
Then, the positive frequency Wightman function of the nonlocal model \eqref{eq:kajuri}
is found as
\begin{equation}\label{eq:G+Kajuri}
G^+(x,x') = \int \frac{d^3 k}{(2 \pi)^3 2\omega} e^{i k_\mu( x^\mu- x'^\mu)},
\end{equation}
which is nothing but the standard Wightman function \eqref{eq:G+local} for the local theory.
Therefore,
the rate of response function is also coincident with Eq.~\eqref{eq:Rdotlocal},
so that
the Unruh effect remains the same as the local case.
This result is consistent with the result
obtained from the Bogoliubov transformation~\cite{Kajuri:2017jmy}.

Finally, it is worth noting that
there is an advantage of our formalism in the sense that
one could discuss the Wightman function \eqref{eq:G+nonlocal} directly
in connection with the Unruh effect.
For the specific Lorentz-invariant nonlocal field theory of \cite{Nicolini:2009dr,Modesto:2017ycz,Kajuri:2017jmy}
\begin{equation}\label{eq:EOMYi}
\square e^{-\ell^2\square/2} \phi_{\rm NL}=0,
\end{equation}
where the Feynman propagator $G_{\rm F}$ is given by
 \begin{equation}\label{eq:FeynmanYi}
G_{\rm F}(x)=\int \frac{d^4 k}{(2\pi)^4}\frac{e^{-\frac{\ell^2}{2}k^2}}{k^2}e^{ik^\mu x_\mu}.
\end{equation}
In Ref.
 \cite{Modesto:2017ycz}, the Feynman propagator \eqref{eq:FeynmanYi} could be nicely expressed as
\begin{equation}\label{eq:Yi}
G_{\rm F}(x)=\int \frac{d^3k}{(2\pi)^3} \left[\frac{1}{2} {\rm erfc}\left(\frac{ \ell}{\sqrt{2}}\omega-\frac{t_{\rm E}}{\sqrt{2}\ell}\right) G^+_k+ \frac{1}{2}{\rm erfc}\left(\frac{ \ell}{\sqrt{2}}\omega+\frac{t_{\rm E}}{\sqrt{2}\ell}\right) G^-
_k\right],
\end{equation}
where $t_{\rm E}$ is the Euclidean time and $G^\pm_k = (2\omega)^{-1}e^{\mp\omega t_{\rm E}\pm i \vec{k}\cdot\vec{x}}$, and the complementary error function is defined as
\begin{equation}
{\rm erfc}(x) = 1-\frac{2}{\sqrt{\pi}}\int^x_0 d\xi e^{-\xi^2}.\notag
\end{equation}
From the fact that the complementary error functions in the Feynman propagator \eqref{eq:Yi}
 become the $\theta$-function
for the limit of $\ell\rightarrow 0$,
it was proposed that the positive frequency Wightman function
 should be written as \cite{Modesto:2017ycz}
\begin{equation}\label{eq:G+Yi}
G^+(x) = \int \frac{d^3 k}{(2 \pi)^3 } G^+_k.
\end{equation}
However, as seen from Eq.~\eqref{eq:Yi}, the complementary error function could not be
factorized if $\ell$ is finite.
Thus, it would be more reasonable to treat the Wightman function  directly from the definition of the Wightman function such as
Eq. \eqref{eq:G+nonlocal} or Eq. \eqref{rel}.
Eventually, our formalism tells us that the Wightman function is the same as that of the local theory
as long as the nonlocal model is Lorentz-invariant, so that the Unruh temperature is accordingly  unchanged.

As mentioned in Ref.~\cite{Modesto:2017ycz}, the authors confirmed that the above model adopted in exploring the Unruh effect gives us the same results of the local field theory. It turns out that the Unruh effect is not modified in
the nonlocal model \eqref{eq:EOMYi} with respect to the local field theory \eqref{eq:EOMlocal}. Consequently, the Unruh
temperature is unchanged.

\section{Conclusion and discussion}
\label{sec:con}

We calculated the Wightman function and the rate of the response function
with respect to the generic nonlocal model \eqref{eq:EOM1} allowing the Lorentz violation
to make the minimal length an invariant scale under any inertial frames.
In this  model,  by using the Unruh-DeWitt detector method, we showed that the Unruh effect could
be corrected by the minimal length generically.
Additionally, taking the limit of the Lorentz-invariant model,
we showed that the Unruh effect always remains unchanged,
which is compatible with the result using the  Bogoliubov transformation in Ref.~\cite{Kajuri:2017jmy}.
Our calculations also explain why the Wightman function read off from the Feynman propagator \eqref{eq:Yi}
is invariant
for the particular Lorentz-invariant nonlocal model \eqref{eq:EOMYi}
in Ref.~\cite{Modesto:2017ycz}.
Therefore, it turned out that the Wightman function and the Unruh effect
remain unchanged if the nonlocal model respects the Lorentz symmetry,
while they should be corrected when the
doubly special relativity is preferred
to make the minimal length an invariant scale.

One might wonder
how the temperature \eqref{eq:tem} could be defined
even though the rate of the response function depends on the proper time.
In the local theory and the Lorentz-invariant nonlocal theory,
the rate of response functions were invariant under the time translation,
which means that
the number of quanta absorbed by the detector per unit proper time $\tau$ is constant,
so that the detector is in a global thermodynamic equilibrium with the scalar field \cite{Birrell:1982ix}.
So, the temperature could be read off
 from the rate of response function
by using the Planck distribution.
However,
in the Lorentz-violating nonlocal theory,
the rate of response function depends on the proper time,
which implies that the system is not in global thermodynamic equilibrium.
So, one should assume that the system  is in a local thermodynamic equilibrium,
where the temperature varies with the spacetime
 but the system is in equilibrium with the neighborhood for each point.
In the local equilibrium,
the scalar field is locally distributed according to the Planck distribution for a certain temperature near a given time when it is observed by the detector.
We could read off the temperature for the Lorentz-violating nonlocal model
approximately in order to figure out how much the thermal temperature deviates
from the standard one.
This issue deserves further attention in this direction.

Finally, we would like to discuss
whether the experiments could be conducted
  to distinguish between Lorentz covariant and Lorentz non-covariant models  at the Planck scale.
In this paper,
the local and nonlocal models 
\eqref{eq:EOMlocal} and \eqref{eq:kajuri} are Lorentz covariant, while the nonlocal model \eqref{eq:EOM1} is 
Lorentz non-covariant.
In our calculations, the models provide the different forms of the Unruh temperature in the sense that
the temperature for the Lorentz covariant models satisfies the standard  Unruh temperature \eqref{eq:T_U}  linearly proportional to the proper acceleration,
whereas the temperature for the Lorentz non-covariant models
generically depends on  the proper acceleration non-linearly as Eq.~\eqref{eq:tem}.
So,
one can distinguish between them
by detecting the Unruh temperatures experimentally.
Actually,
 the large linear acceleration of $2.6\times 10^{22} {\rm cm/s^2}$  is required to produce the Unruh temperature of $1{\rm K}$.
As was suggested in Refs.~\cite{Bell:1982qr, Bell:1986ir}, sufficiently large accelerations could be obtained by employing electrons in a storage ring as an accelerated thermometer.
If one could figure out the behavior of the Unruh temperature with respect to the proper acceleration,
it would be possible to distinguish between the Lorentz non-covariant models
and the Lorentz covariant models experimentally.



\acknowledgments
We would like to thank Myungseok Eune and Sang-Heon Yi for exciting discussions.
W.~Kim, Y.~Gim, and H.~Um were supported by
the National Research Foundation of Korea(NRF) grant funded by the
Korea government(MSIP) (2017R1A2B2006159).
W.~Kim was in part supported by the Sogang University Research Grant (201819009.01).


\bibliographystyle{JHEP}       

\bibliography{references}

\end{document}